\documentclass{aastex61}

\begin{document}

\newpage

\title{Scale invariant cosmology and CMB temperatures as a function of redshifts}
%\title{CMB temperatures as a function of redshifts  from CO molecules and scale invariant cosmological models}

%% Use \author, \affil, plus the \and command to format author and affiliation 
%% information.  If done correctly the peer review system will be able to
%% automatically put the author and affiliation information from the manuscript
%% and save the corresponding author the trouble of entering it by hand.
%%
%% The \affil should be used to document primary affiliations and the
%% \altaffil should be used for secondary affiliations, titles, or email.

%% Authors with the same affiliation can be grouped in a single
\author{Andre Maeder}%\altaffilmark{1}}
\affil{Geneva Observatory, University of Geneva \\
CH-1290 Sauverny, Switzerland \\
andre.maeder@unige.ch}


%\altaffiltext{1}{andre.maeder@unige.ch}

\begin{abstract}
Cosmological models assuming the scale invariance of the macroscopic empty space  show an accelerated expansion, 
without calling for some unknown particles. Several  comparisons between models and observations
(tests on distances, $m-z$ diagram, $\Omega_{\Lambda}$ vs. $\Omega_{\mathrm{m}}$   plot,
 age vs. $H_0$,  $H(z)$ vs. $z$, transition braking-acceleration) 
have indicated an impressive agreement  \citep{Maeder17}.
We pursue  the tests with the CMB temperatures $T_{\mathrm{CMB}}$ as a function of redshifts $z$.
CO molecules in DLA systems
provide the most accurate excitation temperatures $T_{\mathrm{exc}}$ up to $z \approx 2.7$. 
Such data need corrections for local effects, like particle collisions, optical depths, UV radiation, etc. 
We estimate these corrections as a function of the $(CO/H_2)$ ratios 
from   far UV observations of CO molecules in the Galaxy.
%The generally higher  CO column densities of DLA absorbers suggest that these corrections are not overestimated.
The results show that it is not sufficient to apply theoretical collisional corrections to get the proper 
values of $T_{\mathrm{CMB}}$ vs. $z$.  Thus, the agreement  often found with  the standard model may be questioned.
The  $T_{\mathrm{CMB}}(z)$ relation needs 
further careful attention and the same for the scale invariant cosmology in view of its positive tests.
%The   values  of $T_{\mathrm{CMB}}$ are consistent with scale invariant models and
%show their interest.
\end{abstract}




\section{Introduction}  \label{sec:intro}

Unlike theories with modified gravity, the scale invariant models  further explore 
 the invariance properties of space-time, which as emphasized by \citet{Dirac73} play a  fundamental role in physics.
It is well known that the presence of matter in a system tends to  kill  the scale invariance of the physical laws \citep{Feynman63}.
However, the empty space {\emph{at large scales}} may have the property of scale invariance, a property that is present in Maxwell
equations in absence of charge and current. 
\citet{Weyl23}, \citet{Eddi23}, \citet{Dirac73} and \citet{Canu77} have developed a theory which, in addition to the 
 general covariance of General Relativity (GR), also permits (but does not demand)
the invariance to a scale  transformation $ds'= \lambda(t) ds$. 
In this most general framework, the assumption  of  scale invariance of the macroscopic empty space leads to two differential equations
between the cosmological constant $\Lambda$  and   the scale factor $\lambda(t)$.
 Adopting  the postulate of GR   that gravitation universally couples 
to all energy and momentum contributions \citep{Carr92}, we account  for the contribution  from the $\lambda$--derivatives
and obtain new basic  cosmological equations \citep{Maeder17}.
They lead, 
after an initial braking phase, to a general acceleration of the cosmic expansion 
for models with a density parameter $\Omega_{\mathrm{m}} < 1$. 
 %{\bf{These models show some differences with the predictions from GR only when they concern properties 
 %covering  a significant time evolution with respect to the age of the universe}}.

The  properties of these models have been studied  and  detailed tests have been performed \citep{Maeder17},
 in particular on the distances, 
the magnitude--redshift $m-z$ relation, the $\Omega_{\Lambda}$ vs. $\Omega_{\mathrm{m}}$ plot,
the relation between $H_0$ and the age of the Universe,  the expansion rates $H(z)$ vs. redshifts $z$
and  the transition from braking to accelerated expansion. All these tests were very well satisfied without calling for some 
 dark energy. This is why further exploration is necessary.

The $T_{\mathrm{CMB}}(z)$ relation of the  temperatures of the cosmic microwave background (CMB)
 as a function of  redshifts $z$ is a fundamental cosmological test \citep{Peebles93}. 
 Deviations from the standard laws, if they exist, may appear at large enough redshifts. 
 The Sunyaev-Zeld'ovich (SZ) effect  provides some tests  \citep{Luzzi15} for low  $z$.
At such low $z$,  different models predict only small differences
 within the error bars.  \citet{Chluba14} consider that these  tests have a limited applicability.
 The situation is better for the  tests based on molecular absorption in the diffuse intergalactic gas, particular from CO
 lines by  \citet{Srianand08,Noterdaeme10,Noterdaeme11,Noterdaeme17}. The observations of the rotational excitation of the CO
   molecules in Damped Lyman--$\alpha$ (DLA) systems  up to $z \approx 2.7$  have recently been possible \citep{Srianand08}. 
   At present,  remarkably 6 absorption--line systems produced by clouds of diffuse gas have been  detected 
   on the sight lines  of more than 40 000 quasars investigated. 
   The CO molecules emit a spectrum with spectral lines in radio, infrared
and far-ultraviolet  from  rotational, vibrational, and electronic
transitions.  There are several lines observable in the near UV 
and visible spectral domains (due to the redshift). Their simultaneous fitting 
leads to improved determinations of the excitation temperatures. These objects, despite their rareness,
     appear extraordinarily interesting in view of their higher accuracy and the  redshifts they concern.

     In Sect. 2, we study   the  heating and cooling processes intervening in  the excitation of the CO molecules  and try to 
     estimate the global amplitude of these effects from Galactic data.
      In Sect. 3, the excitation  temperatures in DLA systems are discussed and corrections are applied to 
      get the temperatures of the CMB at different $z$.  Comparisons of models and observations are performed  
   in Sect. 4. Sect. 5 gives a  conclusion.

 \section{Contributions to  the excitation of the CO rotational levels} \label{effects}
 
  In addition to the  CMB radiation,    various  physical effects in the diffuse gas may influence the absorption profiles 
  of the fine structure atomic levels and of the rotational levels of 
   molecular  lines, and thus  the determinations of the 
   the excitation temperature $T_{\mathrm{exc}}$ of the  gas.  A  thorough review of the heating and cooling processes
   in the interstellar gas has been given by \citet{Lequeux05}, see also \citet{Wolfire95}. 
   We will concentrate here on the efficient physical effects in 
   molecular regions, with a particular interest on those  potentially able to influence the 
    excitation temperature of the  CO molecules, which have a ground rotational  $J= 1 \rightarrow  0$ transition
    at an energy corresponding to a temperature of  5.54 K. %  with CO lines at 2.6 mm. 
    The significant effects should normally be accounted 
    in order to permit a reliable  CMB temperatures at different redshifts. However,
    corrections for local effects influencing the excitation temperatures  in DLA systems have in general not been applied to the CO observations, (on the six
     observations, only one \citep{Noterdaeme17} has been corrected, see below).
     
   % All these effects need to be very carefully corrected
  % in order to permit a reliable CMB temperature  $T_{\mathrm{CMB}}$. 
   %These are typically  the particle collisions, the pumping by UV radiation,
   % the IR emission from dust, the photon trapping, the other radiation sources. 
    
\begin{figure*}[t!]
\centering
\includegraphics[width=.65\textwidth]{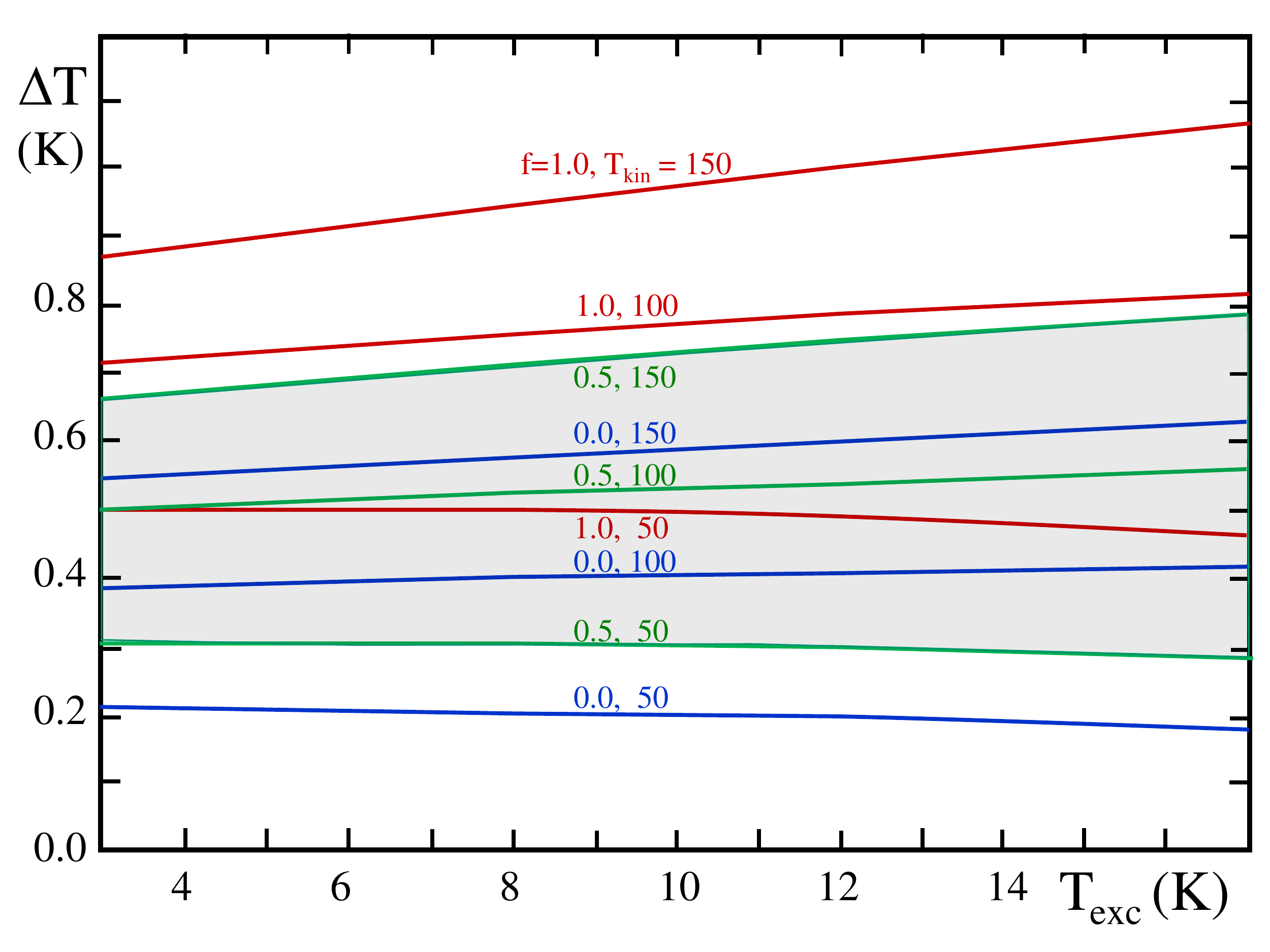}
\caption{The  collisional contributions $\Delta T= T_{\mathrm{exc}} - T_{\mathrm{CMB}}$ in °K for CO molecules ($^{12}$C$^{16}$O)
 as a function  of $T_{\mathrm{exc}}$  for different values  of the molecular
fraction $f$ and of the kinetic  temperature $T_{\mathrm{kin}}$ according to
 expressions by \citet{Sobolev15}. 
A total particle concentration $n=100$ cm$^{-3}$ is assumed, the  corrections $\Delta T$ linearly depend on $n$.   
The gray area corresponds to the regions with  $f=0.5$.     }
\label{Tcorr}
%\end{center}
\end{figure*}

\subsection{Molecular regions: heating  effects influencing the CO excitation}  \label{heating}

There are many processes contributing to the heating of an interstellar molecular region. 
The basic effect is the ejection of an electron  from an atom  
(or a molecule) by an incident photon or a particle,  the electron heating the gas by further collisions 
which rapidly  ($\leq 1$ yr) thermalize the  gas. The main processes are the following ones.
{\it{The heating by low-energy cosmic rays:}} this process is always present due the high penetration of cosmic rays,
 but it is generally insignificant, except in the depths  of molecular clouds where it can even dominate. 
 A cloud model \citep{Noterdaeme17} illustrates the increase of the relative role of cosmic rays with depth in the cloud, 
 where they dominate for relatively low temperatures. 
 {\it{The grain photoelectric emission:}}  it results from the ambient UV flux which removes electrons
 from the grains. The electrons carry a large fraction of the energy of the incident UV photon and
   thermalize the medium by collisions. This process is most efficient in HII and neutral regions, 
   but it may still remain  efficient in the photodissociation regions, intermediate 
   between the HII  and molecular regions. 
   The above mentioned cloud model shows that  this process is still dominant
     in a large part of the cloud where
    CO molecules are present. {\it{The chemical energy of H$_2$ formation:}}  the formation of a  H$_2$ molecule from atomic hydrogen
    is a very exothermal reaction,   yielding energy mainly to the excitation of H$_2$ and the kinetic energy of the gas. 
    This process is present in the outer 
    layers of the the molecular region, though about an order of a magnitude lower than the previous 
    mechanism. {\it{The photon trapping:}} it occurs in regions of  high optical depths, 
    which may largely  reduce 
    the effect of spontaneous emission \citep{Lequeux05}. For example, the de-excitation photons from a CO molecule 
    may be re-absorbed by another CO molecule. This decrease the radiative cooling, so that the interactions with H$_2$
    molecules equilibrate at a higher  temperature \citep{Burgh07}. 
    {\it{The grain--gas thermal exchange:}} it  intervenes due to the collisions between the dust grains
     and  atoms or molecules in the  gas.  
     A high density of the medium favors the process, which  is thus generally unimportant in the diffuse gas, 
   (there, the grains being colder than the gas may  provide some minor cooling). 
 However,  in the depths of giant molecular clouds, the grains may be heated by the far IR radiation,  
 which easily penetrate the clouds. This process 
   dominates in molecular clouds with a concentration $n$ higher than $ 2 \cdot 10^4 $   cm$^{-3}$ \citep{Lequeux05}.
   However,   \citet{Wannier97} have shown that, even in a relatively diffuse gas  with $n$ between  100 
   and 1000 cm$^{-3}$,  the millimeter--wave emission from the dust of a nearby cloud 
   may provide  a significant  specific increase of the  CO-- rotational excitation temperature, (the effect evidently 
   depends on the solid angle presented by the cloud).
    {\it{The UV pumping by  molecules:}} it is  an efficient mechanism only in regions exposed to  a strong 
    UV flux. The  molecules are then  excited rotationally and vibrationally  
    by the absorption of UV photons and the 
    de-excitation transfers some energy to the gas.  This process is often killed by dust extinction \citep{Krotkov80}.
    As CO molecules require screening from UV radiation to exist, this process may  not be dominant for their excitation.

    As to the cooling processes, the cooling by the emission of fine--structure lines
    is mainly due to CII and OI in neutral regions, while in HII regions and deep molecular clouds it becomes insignificant
    \citep{Lequeux05}.  The de-excitation of level $n=2$ of hydrogen, which may be collisionally excited, produces 
    the  Lyman--$\alpha$ line.  In molecular regions, the  main cooling results from the radiation by the fundamental rotational transition 
    of  CO molecules at 2.6 mm. 
    The emission from CI 
    with a first level of excitation at $T=23.4$ K may  also  be significant. 
    The recombination of charged particles on grains, 
    by the  inverse process of the  above mentioned grain photoelectric emission,
     produces a cooling which is rapidly growing with 
    temperature. The grain-gas thermal exchange, as we have seen above, is insignificant in a cold diffuse gas.
    
    Energy losses by induced or stimulated emission do not seem significant processes in the diffuse
    or translucent interstellar medium. Stimulated emission by OH, H$_2$O, NH$_3$, HCN or SiO molecules
    is known to produce the maser effect under certain conditions in dense molecular regions. Even for these relatively easily 
     excited molecules, densities in excess of $10^4$ cm$^{-3}$ are required, 
     {\emph{i.e.}} two orders of magnitude higher than in the diffuse or translucent  interstellar gas.
     To be active, the process also requires a high velocity coherence to avoid Doppler shifts, so that extended
     low density regions are unfavorable.

    The sum of all the  heating and cooling processes leads to a certain equilibrium state in the interstellar medium dominated 
    by H atoms and H$_2$ molecules. In this medium, the particular excitation temperature of the CO molecules  depends 
    on all the above mentioned mechanisms and on their collisional interactions with the main gas components.
  %  is to some extent (but not totally) related to the physical conditions of these main gas components.}}
    % \subsection{Collisional effects for  the CO molecules}  \label{sub:collision}
   \citet{Sobolev15} have recently  studied the population distribution 
   of the  rotational states of the CO molecules ($^{12}$C$^{16}$O) taking into
    account  the CMB radiation  and the collisional excitation by H, H$_2$ and He.  Their results 
     are based on experimentally measured  probabilities of collisional transitions. 
  They provide  the corrections to be applied to  $T_{\mathrm{exc}}$ of 
  CO molecules to get the CMB temperatures. These corrections
   essentially depend on the following four physical factors in the intervening cloud 
    on the sight line of quasars.
 -- 1. The total concentration $n$  of particles in the gas.  
  -- 2.  The hydrogen molecular fraction $f = 2 \, n(H_2)/( \, n(H_2)+n(H))$, where $n(H)$ and $n(H_2)$ are the 
   concentrations of the atomic and molecular hydrogen.  
  --  3. The kinetic temperature $T_{\mathrm{kin}}$ of the gas.    
  -- 4. The  observed excitation temperature $T_{\mathrm{exc}}$ of  the CO molecules.
  Fig. \ref{Tcorr}  illustrates  the corrections   $\Delta T$ derived from
    the analytical expressions by \citet{Sobolev15} as a function of $T_{\mathrm{exc}}$
     due to particle collisions for some typical conditions. These corrections are to be  subtracted from  the excitation 
    temperatures $T_{\mathrm{exc}}$  to get the CMB temperatures  $T_{\mathrm{CMB}}$. 
    The values of  $\Delta T$ depend linearly on the concentration $n$ and 
   are given for different values of $f$ and $T_{\mathrm{kin}}$. 
   We see that they show little variations with the excitation temperatures, 
   while they are more sensitive to the three other parameters. 
   
   Clearly, the above corrections are an important piece of information
   about the corrections to be applied to the  ''observed'' excitation temperature. 
   However, these collisional corrections  do not account for all intervening heating and cooling processes
   influencing the CO excitation,
   %cooling and heating processes act  differently for H$_2$ and CO molecules, so that  equlibrium stages 
  thus  the  effective total corrections  to be applied to   $T_{\mathrm{exc}}$ to get  $T_{\mathrm{CMB}}$ from CO molecules
  may be different from those given in Fig. 1.
   No corrections  have in general   been applied  by the different authors to the observed DLAs with CO lines, except
      for the translucent molecular cloud on the sight line of J0000+0048 (Table 1) which has received a collisional correction
      according to the developments by  \citet{Sobolev15} .
   
  % These corrections are compared  below to the calibrations we can make from data in the Milky Way, 

\begin{figure*}[t!]
\centering
\includegraphics[width=.70\textwidth]{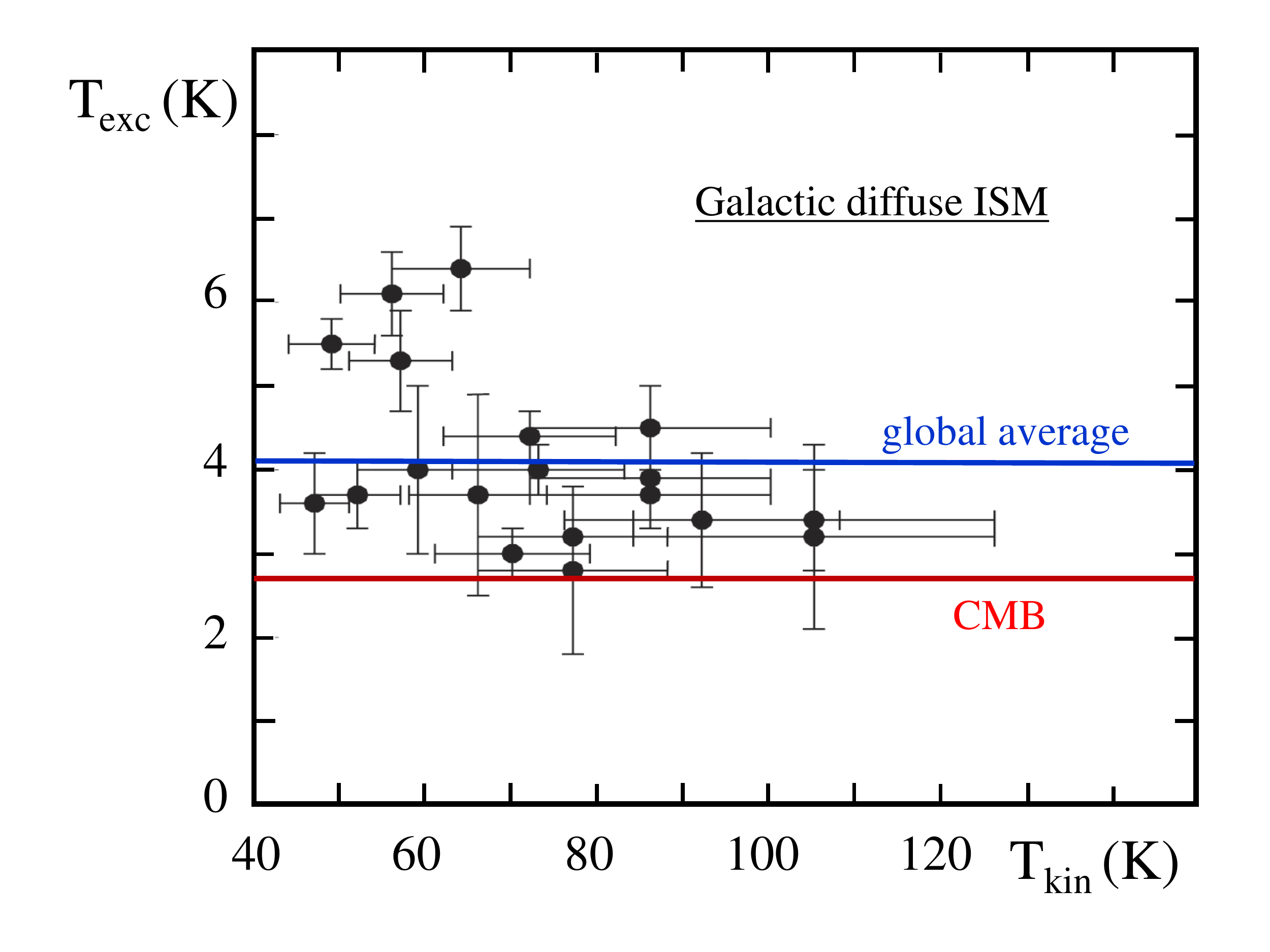}
\caption{The excitation  vs.  kinetic temperatures for CO molecules in the diffuse gas on the sight lines of 23 OB stars  
 in the Milky Way from \citet{Burgh07}, see also \citet{Srianand08}.
The CMB temperature is indicated, as well as the global average of the excitation temperatures.}
\label{Tgal}
\end{figure*}

   \subsection{The excitation temperatures  in the Milky Way and the estimates of the corrections}  \label{MW}
   
\begin{figure*}[t!]
\centering
\includegraphics[width=.75\textwidth]{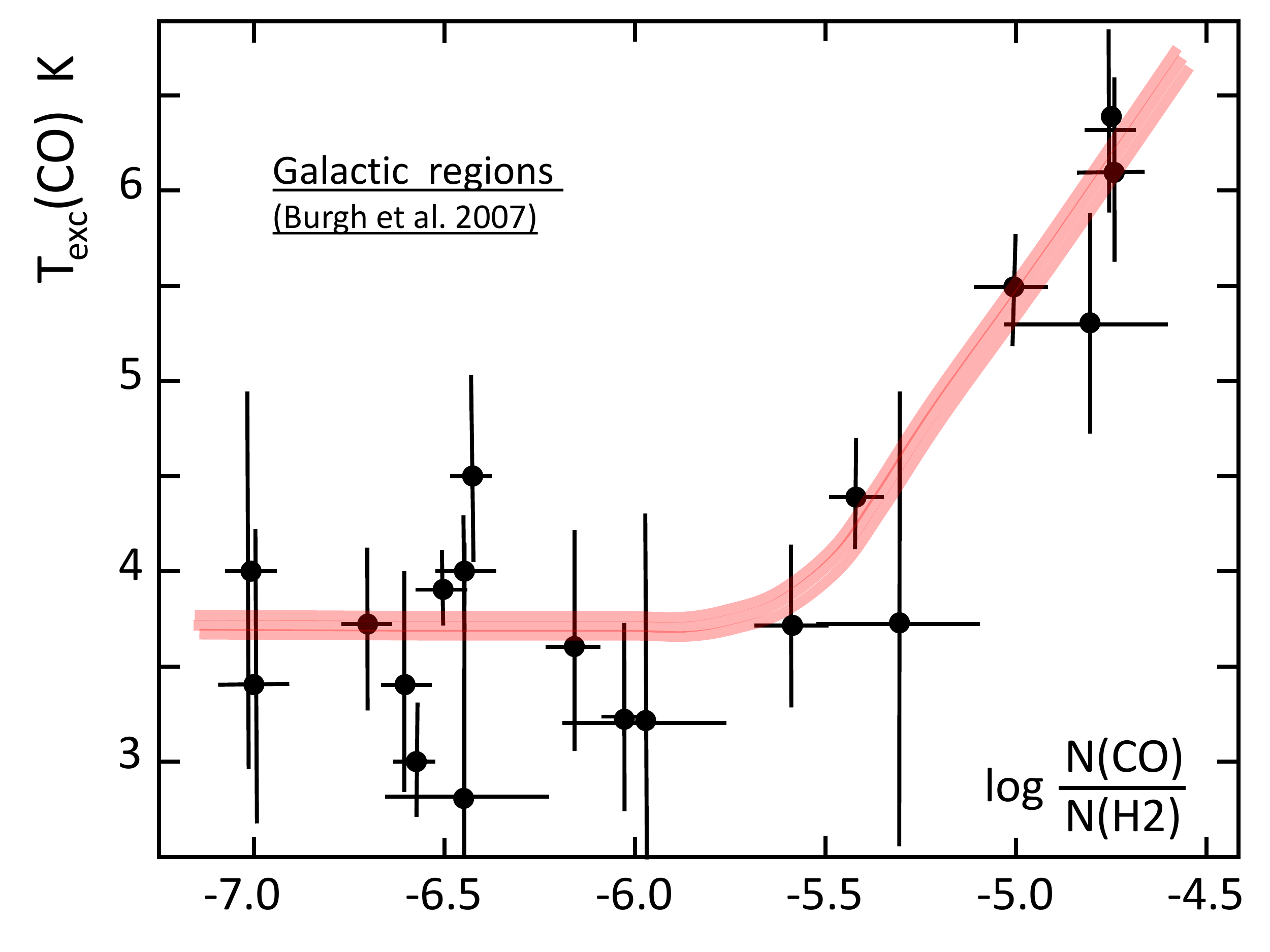}
\caption{The excitation temperature $T_{\mathrm{exc}}$ of CO molecules  vs. $\log \frac{N(CO)}{N(H_2)}$ for diffuse
  Galactic regions  from the  data  by \citet{Burgh07}.  
 A weighted average  of $T_{\mathrm{exc}}=3.69 \pm 0.33$  K is found for  regions with  $\log \frac{N(CO)}{N(H_2)}$    below -5.5.}
\label{NCOH2}
\end{figure*}

CO molecules  produce absorption bands in the UV which permits their detection.   On the basis of  far-UV data from HST STIS and FUSE, \citet{Burgh07} have studied the interrelations 
 between the  physical properties 
 of CO and H$_2$ molecules   in diffuse and translucent regions towards  23 OB stars    in the Milky Way. 
% with an emphasis  on the relation between the $(CO/H_2)$ ratio and the  extinction of the interstellar medium. 
 The average  $T_{\mathrm{kin}}$  of CO molecules ($^{12}$C$^{16}$O) is $74 \pm24$ K, 
with space densities between 20 and 200 cm$^{-3}$ and
   an average $f= 0.22$. (There are 6 sight lines, where $^{13}$CO  could be observed 
   and specific excitation   temperature be determined, but in most cases and even more
   for DLAs, the observation of $^{13}$CO is not achievable at present time).
     The above  ranges covered by the  parameters $n$, $f$ and  $T_{\mathrm{kin}}$,
   as well as the abundance pattern, are rather similar 
    in the diffuse gas of the Milky Way and in DLA systems with CO lines. 
    We note, however,  that 
    the ambient UV flux  generally   appears  higher or equal in  DLA systems  compared to   the Milk Way
     \citep{Ge97,Srianand00,Lima00,Molaro02,Srianand05,Srianand08,Noterdaeme10}.
  As shown by Fig. \ref{Tgal}, the distribution of the CO excitation 
  temperatures $T_{\mathrm{exc}}$ obtained by \citet{Burgh07}
   spans a large range of values above the   local CMB temperature  $T_{\mathrm{CMB}}= 2.726$  K \citep{Fix09}.
   The average of these 
    $T_{\mathrm{exc}}$-values   is  4.095 $\pm 1.01$ K, higher by 1.37 K than the local CMB temperature. 
     \citet{Burgh07} also point out that, for CO column densities from $10^{15}$ cm$^{-2}$ and above,
   %(DLA have  $> 2 \cdot 10^{20}$ cm$^{-2}$)
   the average  $T_{\mathrm{exc}}$ from  CO lines reaches = 5.2 $\pm 1$ K 
    (5.43 K for the  weighted mean),
    {\emph{i.e.}}  2.50 K  (2.7) K higher  than $T_{\mathrm{CMB}}$. 
    For  CO column densities inferior to $10^{15}$ cm$^{-2}$, \citet{Burgh07}
    find an average  $T_{\mathrm{exc}}= 3.6 \pm 0.5$ K,
    (the weighted average   $T_{\mathrm{exc}}$ is 3.69 K, {\emph{ i.e.}} 
    about 0.96 K higher than  $T_{\mathrm{CMB}}$).
    They note that the  increase of $T_{\mathrm{exc}}$ 
      for high CO   (see  also Fig.  \ref{NCOH2}) may result
   from  ''photon trapping'' in regions with a higher optical depth, 
    photons being absorbed by more than one CO molecule so that the radiative cooling is reduced. 
    \citet{Burgh07} also find some relations between the molecular $(CO/H_2)$ ratios 
    and the  extinction properties (color excess $E(B-V)$ and absorption rate
    $A_V/d$ in mag/kpc) for the various diffuse and translucent regions studied.

    Referring to Sect. \ref{heating}, we may  remark that for the above 
     average values of  $T_{\mathrm{kin}}$, $f$ and   $n$
    from 20 to 200 cm$^{-3}$, we have a collisional
   correction between 0.069 K (for $n $ = 20)  and 0.69 K (for $n$ = 200).
    For a mean value  value of the space density $n=100$
   cm$^{-3}$, the typical collisional correction would be 0.35 K. 
   This is  much smaller than what  suggested by the above observations in Fig. \ref{Tgal}. 
   It means that if we account only for the collisional broadening of the lines according  to \citet{Sobolev15},
    we may underestimate by a large amount the total corrections to be
   applied to the observed  $T_{\mathrm{exc}}$ of the CO lines ($^{12}$C$^{16}$O).

%\begin{figure*}[b]
%\centering
%\includegraphics[width=.75\textwidth]{TvsNCO.pdf}
%\caption{The excitation temperature $T_{\mathrm{exc}}$ of CO molecules  vs. $\log N(CO)$ 
 %in the Milky Way data  by \citet{Burgh07}, see also \citet{Noterdaeme10}.  
 %A value of $T_{\mathrm{exc}}=3.7$ is adopted below $\log N{CO}= 15$.}
%\label{TvsNCO}
%\end{figure*} 

     The above mentioned results  by \citet{Burgh07} indicates an effect 
    of the optical thickness of the intervening interstellar gas on the determination of the
    $T_{\mathrm{exc}}$  of CO lines. In a cloud model by \citet{Bolatto13},
    the CO is optically thin for column densities of the CO gas less
     than about $10^{-15}$ cm$^{2}$.  For this limit,  
   the CO gas in DLAs with a low $\log N(CO)$ value
    would  appear to be  generally optically thin.  We note this is not the case for the translucent molecular cloud 
     along the sight line of J0000+0048, even with $log N(CO)= 14.95 \pm 0.05$.  
     Moreover, \citet{Burgh07} in their Fig. 5 show that the Galactic lines 
     of sight with molecular fractions $f$ above  $\sim 0.1$  
    have extinction ratios $(A_V /d)$ of 1 mag/kpc or more. 
    This suggests that the optical depths of the DLA sample
    of Table  1, which when available have $f > 0.2$, may still be significant, 
    if the physical conditions are similar. (I am
    indebted to the referee for these valuable comments). 
  % For a medium of optical thickness $\tau$, the emergent line intensity behaves essentially like $(1-e^{- \tau})$. Thus, 
   For a  thin optically  medium, the emergent line intensity  grows  linearly with 
    the optical thickness times a second term given by
     the populations of the corresponding levels multiplied by the 
    transition probabilities. For high optical depths,  it converges towards the second term. The sample of Galactic data
    by \citet{Burgh07}   cover the low optical depth domain (diffuse gas)
    as well as the transition region to significant optical depths (translucent medium).  The same range is generally present for  DLAs.

    \citet{Burgh07} point out that the  $(CO/H_2)$ 
    ratios are very different for these two regimes, both being   orders of magnitude lower than for dark molecular clouds.
    They find significant relations between the $(CO/H_2)$ ratios
 and  extinction  properties, such as the color excess $E(B-V)$ and 
 the extinction by distance unit $A_V /d$. 
    For the thicker sight lines,  the   $(CO/H_2)$  ratios increase very quickly.
     Thus, in view of these various regimes, we  proceed to individual  temperature corrections
    for each absorption system on the quasar sight lines, using the above CO molecular fraction, that
    correlates with absorption properties as shown by \citet{Burgh07}.
    Fig. \ref{NCOH2} shows the relation between the excitation temperatures of the CO molecules and 
    the $(CO/H_2)$ ratios for the Galactic sample. The 6 regions with $\log N(CO) > 15$ are
    also the 6 regions with $\log \frac{N(CO)}{N(H_2)} > -5.5$ in Fig. \ref{NCOH2}.  For them, we notice a relatively well defined relation 
  with a steep increase   of $T_{\mathrm{exc}}$   with $\log \frac{N(CO)}{N(H_2)}$. This corresponds to the transition region 
  towards the optically thick region  and this  enables  us to perform
    individual corrections in the transition domain.
      For the flat part of the curve corresponding to the  optically thin region below about
    $\log \left(\frac{N(CO)}{N(H_2)}\right) = -5.5 $, the mean temperature  $T_{\mathrm{exc}}$ is 
    3.69 $\pm 0.33$ K, leading to a correction of   $(3.69  - 2.73) =  0.96\pm 0.33 $ K, when 
    compared to the WMAP  determination of the CMB temperature of 2.726 K \citep{Fix09}. For the steep region, the corrections
   will  depend on the observed $(CO/H_2)$ ratios. 
The  question is now:  to what extent can we  apply these last corrections to the DLA sample? This essentially depends on whether 
the physical parameters are similar to those in the Galactic gas. 
  
  \section{The excitation and CMB temperatures in DLA systems}  \label{DLA}

  As mentioned above, from the   40 000 quasars investigated, 
     6  CO absorption--line systems produced by clouds of diffuse or translucent gas on their sight--line have yet  been  detected. 
  In general in the interstellar medium,  carbon is found in different forms from 
  the external to the inner parts of the clouds: ionized, atomic and molecular CO in central
     regions where the shielding from ambient UV radiation is larger. The difficulty is that 
     at the same time the dust extinction is
     generally larger, making  the observations of CO absorption lines more difficult or  preventing them.
   The noticeable successful observations of CO were  performed with the UVES 
   spectrograph on the VLT \citep{Srianand08,Noterdaeme10,Noterdaeme11,Noterdaeme17}.

 \subsection{CO observations and properties in DLA systems}  \label{obs}  
   
 Let us mention these major findings  by order of discovery. Their specific data are given in Table 1, where
  we collect, when available,  the   physical parameters of the diffuse gas in the six DLA absorbers where CO has been observed.
   The first detection  of CO absorption lines  in a DLA system was made by \citet{Srianand08} with  UVES
  at $z=2.418$ towards the quasar J143912+111740.
   The physical parameters given in Table 1  indicate  a  similarity of the physical conditions 
   with those of the diffuse Galactic 
   interstellar material. No collisional correction was applied.
    Thus, the derived value of   $T_{\mathrm{CMB}}$ was taken equal to the observed  $T_{\mathrm{exc}}$,
     which is  0.17 K lower that the theoretically predicted $T_{\mathrm{CMB}}(z)$.

The sub-damped Ly--$\alpha$  system at $z \simeq 2.69$ towards J123714+064759 was studied by \citet{Noterdaeme10}
with the VLT/UVES and X-shooter spectrographs. The mean molecular fraction $f$--value is 0.24, but in some component of the cloud 
it could be close to 1.0. This  absorption region with both atomic and molecular carbon classifies  as a translucent one.
%The derived  physical parameters  are 
%$T_{\mathrm{kin}}=108 $ (+84, -33) K, $f= 0.24 $,  but with  higher values  close to $f=1.0$ in
%he CO--bearing part of the cloud,  the values of $n$ are  in the range 50--60  cm$^{-3}$. 
From the electronic density
they derived, the authors  deduced  that the local UV 
flux in this DLA system is similar to that in the Milky Way.
 No correction to $T_{\mathrm{exc}}$ was  applied to get $T_{\mathrm{CMB}}$, despite the fact that this translucent region 
 is unlikely optically thin.

Three new CO absorption-line systems were observed by \citet{Noterdaeme11} with the same instrument and reduction process as before. 
A system at  $z=1.774$  was observed on the sight line of 
J104705+205734.  Two systems  were also  
observed, one  at  $z=1.729$ towards J085726+185524, and another one at $z=2.038$ towards J170542+354430. 
These two systems   are hardly resolved, but best--fit models  nevertheless provide the
excitation temperatures. No corrections were applied to these three systems.

\begin{table*}[t!]  
\vspace*{0mm}  \label{data1}
 \caption{Physical parameters in the DLA systems with CO absorption lines and determinations of the excitation temperatures.
  The last column gives the references.  } 
\begin{center}  
\scriptsize
\begin{tabular}{ccccccccccc}
  Quasar   &  $z$  & $T_{\mathrm{kin}}$ & $f$ & $n$ & $T_{\mathrm{exc}}$(CO) &log N(CO) & log N(H$_2$) &
  log $\frac{CO}{H_2}$ &ref.& \\
      &    &  (K) & $f$ & (cm$^{-3}$) & (K) &  & & & & \\
   \hline
 &   &   &   \\
J0000+0048         & 2.5255 &   50                   &0.46$^{+0.07}_{-0.07}$& 80    &9.9$^{+0.7}_{-0.6}$  
&14.95$\pm0.05$ & 20.43 $\pm0.02$ & -5.48$\pm0.05$   &  4 &\\
J085726+185524 & 1.7293 &    --                     &           --                         &   --    & 7.5$^{+1.6}_{-1.2}$ & -- &  -- & -- & 3 &\\
J104705+205734 & 1.7738 &     --                    &           --                        &  --     & 7.8$^{+0.7}_{-0.6}$ & 14.74 $\pm0.07$ & -- & -- & 3 &\\    
J123714+064759 & 2.6896 &108$^{+84}_{-33}$&0.24 (1.0)                & 50-60& 
10.5$^{+0.8}_{-0.6}$ & 14.17$\pm0.09$& 19.21 $^{+0.13}_{-.012}$ & -5.04$\pm0.16$ & 2&\\  
J143912+111740 & 2.4184 &  105                 &0.27$^{+0.10}_{-0.08}$&  45-62&9.15$^{+0.7}_{-0.7}$ &
13.89$\pm0.02$ & 19.38 $\pm0.10$ & -5.49$\pm0.10$ &1&\\ 
J170542+354340 & 2.0377 &    --                     &           --                         &   --  & 8.6 $^{+1.1}_{-1.0}$ &  -- &--& --   & 3&\\
\hline
\end{tabular}
\end{center}
\vspace{-1mm}
{References: 1.-- \citet{Srianand08}.  2.-- \citet{Noterdaeme10}. 3.-- \citet{Noterdaeme11}.  4 .--\citet{Noterdaeme17}.}
\end{table*}%\vspace*{2mm}
\normalsize

   A molecular cloud classified as a DLA at $z \approx 2.53$ along the sight line of J0000+0048 has been recently
   studied in great details by \citet{Noterdaeme17}.
  %The kinetic temperature in the inner cloud  is $T_{\mathrm{kin}} =50$ K, with values of 70 to 80 K in most of the cloud.
 % and the  H particle density $n=80$ cm$^{-3}$,  ranging  between 40 and 100   cm$^{-3}$.
  The derived  molecular fraction is $f=0.46\pm 0.07$, which is the highest average $f$ observed in a high-$z$ intervening system.
   The properties of this translucent regions compare well with those of the Galactic Perseus complex 
   \citep{Noterdaeme17}.
   The collisional correction derived by the authors amounts to 0.3 K on the basis of the data by \citet{Sobolev15}. 
   Applied to the observed   $T_{\mathrm{exc}}=9.9$ K (+0.7,-0.6),
   it leads   to a value $T_{\mathrm{CMB}}$, in agreement  with the standard  
   value for this redshift.

%\begin{figure*}[t!]
%\centering
%\includegraphics[width=.65\textwidth]{NCO.pdf}
%\caption{Relation between the  excitation temperature $T_{\mathrm{exc}}$
%and the column densities $N(CO)$  of the Galactic  diffuse gas from the data by \citet{Burgh07}. Bins of four successive points
%in $N(CO)$ are performed. The error bars gives the mean square root. The vertical red broken line indicates  the
%mean $N(CO)$-value of the DLA sample of Table 1.  }
%\label{NCO}
%\end{figure*}
   
   \subsection{Applications of the Galactic temperature corrections} \label{apply}

There is evidently a great variety of physical conditions in DLA systems. 
Some also have an UV radiation field up to 10-100 times the Galactic one, in that case the molecular fraction is very low \citep{Reimers03}
and  such systems show no CO molecular absorption.
 %For example,  for an absorption  system at $z=1.973$, \citet{Ge97} have proposed a radiation field 
%higher by an order of a magnitude than in the Milky Way, see also \citet{Lima00}
However, the systems with C and CO molecules typically show physical conditions and  UV radiation fields close to those 
of the Galactic diffuse gas. To some extent, the presence of CO molecules imposes a limited domain
 in the space of the physical parameters.
\citet{Molaro02} detect a Galactic type  UV radiation field  in  a DLA system at $z=3.025$ with various  C absorption lines. 
\citet{Srianand08} conclude that the physical conditions in the object at $z=2.418$ (with the first discovered  CO lines)
are similar to those in the diffuse Galactic clouds. For the second discovery, \citet{Noterdaeme10} find a 
molecular fraction $f$ typical of the Galactic conditions. They also point out the similarity of the abundance patterns.
 \citet{Noterdaeme17} also  find physical properties (density, molecular fraction, UV radiation,...) 
in the Perseus-like system at $z=2.5255$ very similar to the  conditions 
 observed in the Milky Way.
All these comparisons support the application of the  temperature corrections based on the Galactic diffuse gas to the  
excitation temperature of CO molecules in DLA systems.

A study of of the physical conditions in  33 DLA systems  with significant molecular absorption lines of H$_2$ and lines  of carbon
in various excitation states 
has been performed by \citet{Srianand05} on the basis of a sample collected 
by \citet{Ledoux03}. The mean kinetic temperature is 153 $(\pm 78)$ K, significantly  higher than the Galactic one (74 $(\pm 24)$ K.
The mean H$_2$ molecular fraction $f$ found by \citet{Ledoux03} is smaller than 0.1. This value is
 in fact an average over the whole line of sight, the actual $f$ in the individual DLA components may be much larger,
  as emphasized by \citet{Srianand05}.
%Thus, it is likely  not smaller than the average Galactic $f$ value of 0.22. 
 The mean particle concentration  is
$n= 78$ cm$^{-3}$ quite in the observed Galactic range of 20-200 cm$^{-3}$. As to the ambient UV radiation field,
\citet{Srianand05} conclude that it is of the order or slightly higher  than the mean UV radiation field in the Galactic 
interstellar material.  
These results also suggest that the local excitation processes  of molecular lines are equal or slightly higher in DLA clouds
than in the Galactic interstellar medium.

Table 2  gives the various determinations of the CMB temperatures. The $T_{\mathrm{CMB}}$ given by the different authors are
given in column 3. Except for J0000+0048  they are identical to the determined excitation temperatures
      (see Table 1). The corrections determined from the Galactic data in Fig. \ref{NCOH2} are given in column 4. 
      The corrections amounts to   0.96 $\pm0.33$ K for objects belonging the flat part of the curve 
       below $\log (CO/H_2)= -5.5$. For objects with  $\log (CO/H_2)$ above this last value, 
       the corrections are determined individually from the red curve in Fig. \ref{NCOH2}
     with the same error bar  of  $\pm0.33$ K, despite the fact that  there the scatter may appear lower.
     The  $T_{\mathrm{CMB}}$--values obtained with these corrections are given
     in column 5 of the Table 2. To enable further comparisons, the theoretical values corresponding to the observed redshifts 
     are given in columns 6 and 7 for the standard and scale invariant cosmology respectively according to Sect. \ref{cosmo}.

\begin{table*}[t!]  
\vspace*{0mm}  \label{data2}
 \caption{The various determinations of the CMB temperatures. 
  Column 3 gives $T_{\mathrm{CMB}}$ 
  derived by the various authors quoted in Table 1, (no correction was applied except for  J0000+0048); the error bars are the same as for $T_{\mathrm{exc}}$(CO) in column 6 of Table 1.  Column 4  shows the temperature corrections 
   based on the Galactic data by \citet{Burgh07} (Sect. \ref{MW}).  Column 5 gives the values of  
$T_{\mathrm{CMB}}$ obtained with the corrections of column 4.  Columns 6 and 7 give the predicted $T_{\mathrm{CMB}}$ 
for the standard and scale invariant theory respectively (see Sect. \ref{cosmo}). } 
\begin{center}  
\scriptsize
\begin{tabular}{ccccccc}
  Quasar   &  $z$  &  Obs. $T_{\mathrm{CMB}}$ & T-corrections & Obs. $T_{\mathrm{CMB}}$ &
  $T_{\mathrm{CMB}}(z)$ & $T_{\mathrm{CMB}}$(z) \\
      &   &  ref.1,2,3,4 & from Sect. \ref{MW}&   corrected  & Std. th. &  Sc. inv.   \\
   
   \hline
 &   &   &   \\
J0000+0048         & 2.5255 & 9.6$^{+0.7}_{-0.6}$     &  2.57 $\pm0.33$ &  7.33 $^{+0.77}_{-0.68}$        &  9.61   & 8.27      \\
J085726+185524 & 1.7293 & 7.5$^{+1.6}_{-1.2}$      &  0.96 $\pm0.33$ &  6.54$^{+1.63}_{-1.24}$         &  7.44  & 6.53      \\
J104705+205734 & 1.7738 & 7.8$^{+0.7}_{-0.6}$      &  0.96 $\pm0.33$ &  6.84$^{+0.77}_{-0.68}$         &  7.56  &  6.62       \\    
J123714+064759 & 2.6896 &10.5$^{+0.8}_{-0.6}$     &  1.32 $\pm0.33$ &  9.18 $^{+0.87}_{-0.68}$        &  10.06  & 8.63      \\  
J143912+111740 & 2.4184 &9.15$^{+0.7}_{-0.7}$     &  1.30 $\pm0.33$ &  7.85$^{+0.77}_{-0.77}$         &   9.32  &   8.03     \\ 
J170542+354340 & 2.0377 &8.6 $^{+1.1}_{-1.0}$      &  0.96  $\pm0.33$&  7.64$^{+1.15}_{-1.05}$         &   8.28  &   7.20   \\
\hline
\end{tabular}
\end{center}
\end{table*}%\vspace*{2mm}
\normalsize

There is another concern, related to  the slightly higher values of $T_{\mathrm{exc}}$  in the  DLA sample,
from 7.5  to 10.5 K (Table 1),   compared to  2.8 to 6.4 K  in the Galaxy (Fig. \ref{Tgal}).
%We may wonder whether, even for similar gas parameters, the corrections would be exactly the same
 %in the Galactic gas and in the DLAs.}}
% It could be argued that the UV pumping could act differently between the DLA's and the Galaxy.
 %In view of the differences between the energies of the rotational levels considered, such effects are likely small. 
  The values of  $T_{\mathrm{exc}}$ are determined by assuming a Boltzmann distribution  of the populations  $N(CO,J)$ of the $J$
  rotational levels, which are fitted by a single value $T_{\mathrm{exc}}$. This value
  is given by the slope of a plot $\log N(CO,J)/g_J$ vs. the energy $E_J$ of the  $J$--level considered ($g_J$ being the statistical weight).
   At least four values of $E_J$ have
   been considered by the authors quoted in Table 1
  up to a value corresponding to 33 K. For J0000+0048, even a value at about 55 K has been used. 
  In view of the  broad ranges of level energies involved in the plot, a difference  between $T_{\mathrm{exc}}=4$ K and 9 K
  only has a very limited effect on the $T$--corrections. 
Moreover, Fig. \ref{Tcorr} shows that  the collisional corrections slightly increase with   $T_{\mathrm{exc}}$.
% from 4   to 9 K, the  
%$T$--corrections for collisional effects increase by  about 0.22 K for  typical values $n=100$ cm$^{-3}$,  
%$T_{\mathrm{kin}}=100$ K and  $f=0.5$. For the same 
% $T_{\mathrm{kin}}$,   with $n=50$ cm$^{-3}$ and $f=0.25$, the correction increases by about 0.1 K. 
 %Thus, as far as the collisional contribution is  concerned, it should be {\emph{slightly larger}} 
 % $T_{\mathrm{exc}}=9$ K in DLA systems than at 4 K 
 %in the Milky Way.
 % Thus, we conclude that the corrections derived in Sect. \ref{sub:MW} and applied in column 8 of Table 1
 These facts tend to indicate that the corrections we apply
 are  not overestimated, especially more than the UV flux is equal or higher in DLA  systems than in the Milky Way.

\begin{table*}[t!]  
\vspace*{0mm} 
 \caption{Predicted  $T_{\mathrm{CMB}}$ as a function of redshift $z$ in the standard case and in the scale invariant case
 for $k=0$ and $\Omega_{\mathrm{m}}=0.30$.
 Column 1 gives the time $t$ in a scale where $t_0= 1$, column 4 gives  $\lambda^{-1/2}$.}
\begin{center} 
\scriptsize
\begin{tabular}{ccccc}
  $z$    &  $t$      & $T_{\mathrm{CMB}}$ & $\lambda^{-1/2}$ &  $T_{\mathrm{CMB}}$ \\
           &             &        standard              &                              &  scale invariant           \\                               
           \hline
%          &              &                                   &   \\
 0       & 1           &       2.726                    &             1               &       2.726                  \\
%0.1     & 0.96790 &       2.999                    &     0.98382            &        2.950                  \\
0.2     & 0.94073 &       3.271                    &     0.96991            &        3.173                  \\
0.4     & 0.89735 &       3.816                    &     0.94728            &         3.615                \\
0.6     &0.86441  &       4.362                    &     0.92973            &         4.055                \\
%0.8     & 0.83866 &       4.907                    &     0.91578            &        4.494                \\
1.0     & 0.81807 &       5.452                    &     0.90447             &        4.931                \\
1.5     & 0.78138 &       6.815                    &     0.88396             &         6.024               \\
2.0     &0.75753  &       8.178                    &      0.87035            &          7.118              \\
2.5     &0.74102  &       9.541                    &      0.86082            &          8.213              \\
3.0     &0.72905  &      10.904                   &      0.85384            &          9.310              \\
%3.5     &0.72006  &       12.267                  &      0.84856            &          10.409            \\
4.0     &0.71309  &       13.630                  &      0.84444            &           11.510           \\
10.0   &0.68341  &       29.986                  &      0.82669            &            24.789          \\
$10^3$& 0.66945&   2.728 $10^3$         &   0.81820                   &   2.232 $10^3$      \\ 
\hline
\end{tabular}
\end{center}
\end{table*}
 \normalsize 
 
 \section{Comparisons with cosmological models}   \label{cosmo}
 
\subsection{The classical model}  \label{sub:classical}
\begin{figure*}[t!]
\centering
\includegraphics[width=.80\textwidth]{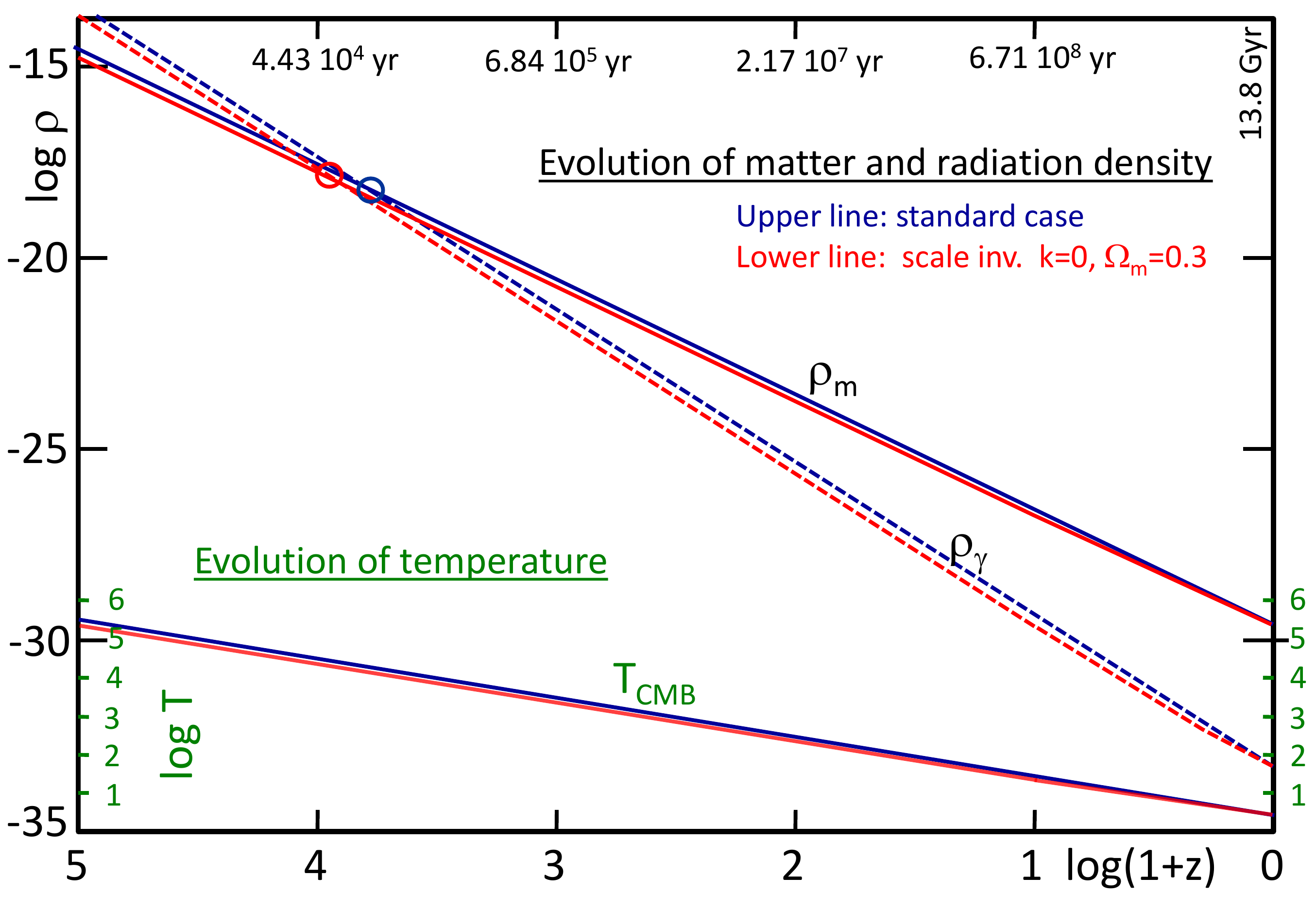}
\caption{Past variations  of $T_{\mathrm{CMB}}$, the radiation density $\rho_{\gamma}$ and the matter density $\rho_{\mathrm{m}}$ 
 as a function of redshifts  for the standard and scale invariant models. Age indications are given. The present values 
 are $\log \rho_{\gamma}=-33.333$, $\log \rho_{\mathrm{m}}=-29.559$,
  $\log T_{\mathrm{CMB}}=0.435$.  For $\log (1+z) > 2 $
the separation of the curves is very close to a limiting constant. 
For $T_{\mathrm{CMB}}$, this limiting constant  amounts to 0.087
dex,  to 0.349 dex for $\rho_{\gamma}$ and to 0.174 dex for $\rho_{\mathrm{m}}$.  Small circles mark the crossing points
of classical and scale invariant models.}
\label{early}
\end{figure*}

Since matter and radiation are decoupled in the present era of the Universe, energy
conservation implies   
 for the radiation energy density $\rho_{\gamma} \, R^4 = const$. In turn, since   $\rho_{\gamma} \sim T^4$,
we have  $T \, R = const$. With  $R_0/R= 1+z$, this gives  the classical relation for the CMB temperature 
 $T_{\mathrm{CMB}} (z)$ law, see for example  \citet{Peebles93},
\begin{equation}
T_{\mathrm{CMB}} (z) \, = \, T_{\mathrm{CMB}} (0) \, (1 + z)   \, .  
\label{std}
\end{equation}
\noindent
The blue line in Fig. \ref{early} illustrates the past variations of $T(z)$, as well as of the matter
 and radiation densities $\varrho_{\mathrm{m}}(z)$
and $\varrho_{\gamma}(z)$ respectively.
These relations  are not sensitive to the shape $R(t)$ of the expansion function (only for the classical case),  as are the tests based on distances
such as the $m-z$ diagram \citep{Riess98,Perl99},  or as the past expansion rates $H(z)$ vs. redshifts  
\citep{Farooq13}. 
%Relation (\ref{std})  is specifically testing the global conservation of photons number per covolume unit,
 %or more generally the energy conservation (in absence of photon creation). }}
%As mentioned in the introduction, Relation has been studied on the basis of the Sunyaev-Zel'dovich (SZ) effect at low $z$, $z  < 0.6$,  
%  see for example \citet{Luzzi15} and \citet{Noterdaeme11}. % The conclusions were generally supporting the above classical relation. 
%Atomic and molecular absorption lines  produced by the  diffuse  gas in Damped Lyman $\alpha$ system on the sight line
%of quasars have provided a most interesting information on the CMB temperature higher redshifts.
% Observations  provided by CO molecules go to large enough  redshifts,   of about $z=2$ to 3, to allow new interesting cosmological tests. 

\subsection{The scale invariant model}  \label{scale}

\begin{figure*}[t!]
\centering
\includegraphics[width=.99\textwidth]{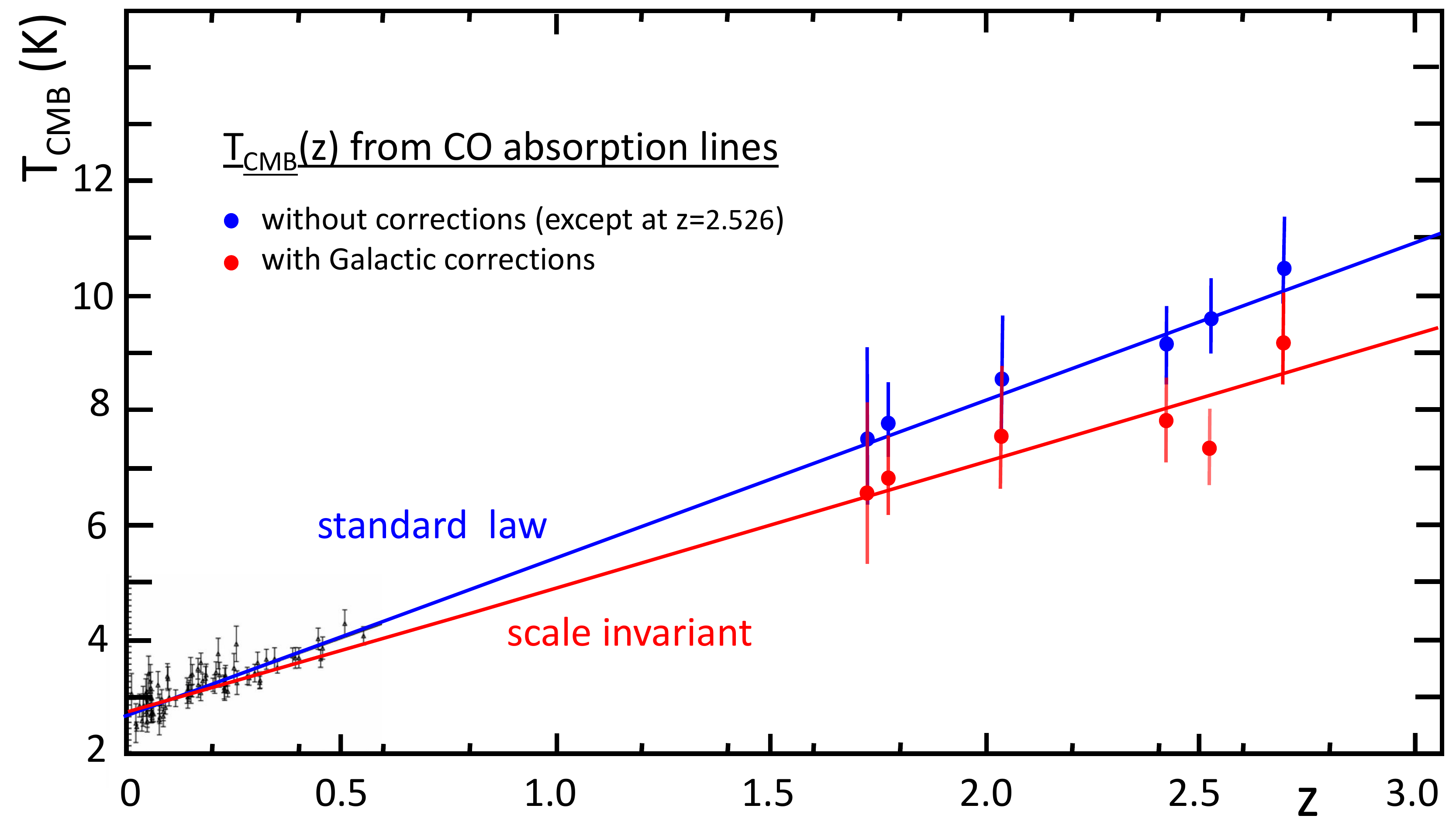}
\caption{The temperature of the CMB vs. redshifts. The blue line gives the standard relation \citep{Peebles93} and
 the red line for the scale invariant models with $\Omega_{\mathrm{m}}=0.30$ \citep{Maeder17}. These relations
 are given by Eqs. \ref{std}  and \ref{sci} respectively.
The blue points show   $T_{\mathrm{CMB}}$  \citep{Srianand08,Noterdaeme10,Noterdaeme11,Noterdaeme17}, which are   identical to $T_{\mathrm{exc}}$(CO), except for the value at $z \simeq 2.526$.  The red points give the values $T_{\mathrm{CMB}}$
 derived with the empirical corrections from Sect. \ref{MW}, see Fig. \ref{NCOH2} based on the data by \citet{Burgh07}. Indicative
points below $z=0.6$ are   reproduced from the data  on the SZ effect by \citet{Luzzi15}. }
%The green  open squares are based on the fine structure of atomic carbon, at $z \approx 1.8$
%by \citet{Cui05}, at $z \approx 2.0$  by \citet{Ge97}, at $z \approx 2.3$ by \citet{Srianand00} and at $z \approx 3.0$ by  
%citet{Molaro02}. }
\label{Tvsz}
\end{figure*}

 Several basic properties of the scale invariant cosmology, that accounts for the scale invariance of the empty space, were
 studied recently \citep{Maeder17}. The numerical solutions of the cosmological models were presented. Comparisons
 with observations were performed for several major tests, with positive results. 
 From the cosmological equations, 
 we got the following equation of conservation,
\begin{equation}
\frac{d(\varrho R^3)}{dR} + 3 \, p R^2+ (\varrho+3\, p) \frac{R^3}{\lambda} \frac{d \lambda}{dR} = 0 \, ,
\label{conserv}
\end{equation}
\noindent
which for a constant $\lambda$ evidently gives the classical expression.
 From   the equation of state in the general form
$P \, = \, w \,  \varrho $   with $c^2$ =1 ($w$ being  taken here as a constant)
and after integration, we obtained
\begin{equation}
\varrho \, R^{3(w+1)}  \,  \lambda ^{(3w+1)} \,=  const.  
\label{3w}
\end{equation}
\noindent
It implies that the curvature of space associated to a distribution of mass--energy  has some dependence on $\lambda$ and thus on time,
since the scale factor $\lambda$ is equal to  $t_0/t $ \citep{Maeder17}.
For radiation density $\varrho_{\gamma}$,  
we have $w=1/3$ and thus 
$\varrho_{\gamma} \,  R^4 \, \lambda^2 \,=const.$  This leads to $T  \,  R \, \lambda^{1/2} \, = const.$
and the temperature of the CMB radiation behaves like
\begin{equation}
T_{\mathrm{CMB}} (z) \, = \, T_{\mathrm{CMB}} (0) \, (1 + z)  ({t}/{t_0})^{1/2} \, ,
\label{sci}
\end{equation}
\noindent
  The domain of $\lambda$--values  depends on 
the density parameter $\Omega_{\mathrm{m}}$. For an empty model with $\Omega_{\mathrm{m}}=0$, 
one would have $t/t_0 = 0$  at the origin  and
thus $\lambda$ would enormously  vary,  from $\infty$ to  1 at present. If matter--energy  is present in the Universe,
 the effects of scale invariance
tend to disappear \citep{Feynman63}. For $\Omega_{\mathrm{m}}=0.30$, the whole range is limited 
from  $\lambda=1.494$ at the origin to 1 at present.% For $\Omega_{\mathrm{m}}  \geq 1.0$, $\lambda$ is always equal to 1.
The value $\Omega_{\mathrm{m}} = 1.0$ appears as the limit above which scale 
invariance  is absent from cosmological models.

Fig. \ref{early} shows the past evolution of $\rho_{\mathrm{m}}$, $\rho_{\gamma}$ and $T_{\mathrm{CMB}}$
as a function of redshifts. We note the very small separations between the standard and scale invariant cases.
The curves start separating at low $z$ and then  from  $ z \approx  10^2$ onwards their separations remain about constant,
 because the value of $t/t_0$ does not change significantly in the early explosive phases. The highest separation amounts to    0.087 dex 
 for $ T_{\mathrm{CMB}}$, 0.174 dex for $\rho_{\mathrm{m}}$ and 0.349 dex for $\rho_{\gamma}$. Thus,  the differences 
 of the two  models  in the early phases of the Universe are very limited.
Table 3 gives the  CMB temperatures    $ T_{\mathrm{CMB}}$ both in the standard  and in the scale invariant case (for 
$\Omega_{\mathrm{m}}=0.30$)  for some values of $z$. 
%the value of $\lambda^{-1/2}$ is also indicated.
%We see the slightly growing differences between the  two $ T_{\mathrm{CMB}}$ predictions.
  The temperature in the scale invariant model  is lower by about 1 K at $z=2$ than in the standard case, corresponding  to
-0.060 dex. This shows  that a relatively high accuracy of the observations is necessary for a  significant analysis.

\subsection{Comparison of models and observations}

   Fig. \ref{Tvsz} shows the variations of  $ T_{\mathrm{CMB}}$
for both the standard and scale invariant cosmological models  as a function of redshift $z$.
The theoretical data are compared to the temperatures derived from the CO molecules as given in Table 2.
%The four green open squares, which show
%values from the  fine structure lines of atomic carbon,  are more  in favor in the standard case. These data
%have received some corrections by their  authors, however there is no check of  the validity of these corrections.
The blue points represent  the values of $T_{\mathrm{CMB}}$  from CO molecular lines in DLA systems
as given by the authors (see Table 1). The values  $T_{\mathrm{CMB}}$ were taken identical to  $T_{\mathrm{exc}}$, 
(the point at $z \simeq 2.53$ has been corrected for collisions).  These  blue points are supporting the standard case.
 The red points in Fig. \ref{Tvsz} give
  the values of  $ T_{\mathrm{CMB}}$  obtained with  the  Galactic corrections as established  in Sect. \ref{MW}, see Table 2.
   These last results appear to favor 
the scale invariant  cosmological  models. However, one must still be careful in the interpretation of the results, in view
of the small  number of  CO observations, which already represent a most remarkable achievement. Coming instrumental developments
 of high resolution spectrographs will certainly enlarge the sample of these key observations.

\section{Conclusions}

The corrections to be brought to the $ T_{\mathrm{exc}}$
of CO molecules seen in absorption on the sight lines of quasars in order to get $T_{\mathrm{CMB}}(z)$  is a complex problem.
Nevertheless, one can  safely conclude that the support 
  generally given to the standard model by these CO observations  may be questioned.
The present results   suggest that it is not sufficient  to assume that $T_{\mathrm{CMB}}(z) = T_{\mathrm{exc}}$(CO)  
at the observed redshifts, or to only apply the collisional corrections.
%The   $T_{\mathrm{CMB }}$  vs. $z$  relation at high redshifts deserves further  careful attention.
  The assumption of the scale invariance of the macroscopic empty space is, at least for now,
  not contradicted by these observations and it  deserves further attention.

\allauthors

\listofchanges

\end{document}